\documentstyle[11pt,newpasp,twoside,epsf]{article}
\def\edcomment#1{\iffalse\marginpar{\raggedright\sl#1\/}\else\relax\fi}
\reversemarginpar

\begin{document}
\title{Galactic Mergers with Supermassive Black Holes}
 \author{Fidel Cruz$^{1,2}$ \& David Merritt$^1$}
\affil{$^1$Department of Physics and Astronomy, Rutgers University, NJ, USA}
\affil{$^2$Instituto de Astronom\'{\i}a, UNAM, Ensenada, B.C., M\'exico}

\begin{abstract}
We present the  results of $N$-body simulations of the 
accretion of high-density dwarf galaxies by 
low-density giant galaxies. Both galaxies contain 
power-law central density cusps and point masses
representing  supermassive black holes; the ratio
of galaxy masses is $3:1$.
The cusp of the dwarf galaxy is always disrupted during the merger, 
leading to a remnant with a weak power law in the intrinsic
density and a ``core'' in the projected density. Removing both black 
holes from the giant and dwarf galaxies allows the dwarf galaxy 
to remain intact and leads to a remnant with a high central density, 
contrary to what is observed. Our results support the hypothesis 
than the persistence of low-density cores in giant galaxies 
following mergers is a consequence of the existence of 
supermassive central black holes.
\end{abstract}

\section{Introduction}

Bright elliptical galaxies and bulges of spiral galaxies tend to
be less dense than faint galaxies. The luminosity density in the 
central regions of these galaxies rises approximately as a power law 
at the smallest observable radii, $\rho\sim r^{-\gamma}$. Faint 
galaxies ($M_V\ga -20$) have $1.5 \la \gamma \la 2.5$ while bright 
galaxies have $ 0 \la \gamma\la 1.5$ 
(Ferrarese et al. 1994; Gebhardt et al. 1996). 
There are some interesting questions relating to these observed facts: 
(1) How did the cusps form? 
(2) How are the weak cusps in  bright
galaxies maintained in spite of mergers with dense low-luminosity
galaxies? 
If the small galaxy survived such a merger with its central 
regions intact, this would create a new high-density core in the center 
of the big galaxy and destroy the observed correlation 
between $M_V$ and $\gamma$.
Strong cusps form naturally in stellar systems where the
black hole grows on time scales long compared with the crossing time
(Peebles 1972; Quinlan et al. 1995). 
Milosavljevi\'c \& Merritt (2001) presented 
simulations where the initial galaxies have steep central
density cusps (like those in low-luminosity galaxies) 
and central black holes. The  central
density cusp in the final remnant had a slope $\rho \sim r^{-1}$
after the binary black hole had ejected a mass of order its own
mass from the pre-existing nucleus. This result can 
explain how shallow cusps form. In this work, we describe a set of 
merger simulations that address the question
of how a shallow cusp can {\it survive} the accretion of a small dense galaxy.
Our work extends that of Merritt \& Cruz (2001; hereafter Paper I) 
who first presented merger simulations of galaxies containing central 
black holes and density cusps.

\section{Method}

Initial galaxies were generated from Dehnen's (1993) law,
\begin{equation}
\rho(r) = {(3-\gamma)M\over 4\pi a^3}\left({r\over a}\right)^{-\gamma}\left(1+{r\over a}\right)^{\gamma-4}
\end{equation}
which has a power-law central density dependence, $\rho \propto r^{-\gamma}$.
Our primary (massive) galaxies had $\gamma=1$, characteristic of the shallow 
cusps in bright galaxies, and our secondary (dwarf) galaxies had $\gamma=2$, 
characteristic of the steep cusps of dwarf galaxies. Subscripts 1 and 2
will henceforth refer to the primary and secondary galaxies respectively.
Initial particle velocities were assigned from  
an isotropic distribution function that accounts for the central point 
mass representing the black hole (Tremaine et al. 1994).   
Each black hole was assigned a mass 2$\times 10^{-3}$ times 
that of the parent galaxy. This is consistent with the best estimated 
value of $\sim 0.0012$ for the mean ratio of black hole mass to 
luminous galaxy mass in the local universe (Merritt \& Ferrarese 2001).

The mass $M_1$ and length scale $a$ of the primary galaxy were
set to unity. 
We chose $M_1/M_2=3$; this mass ratio complements the more extreme
ratio of $M_1/M_2=10$ adopted in Paper I.
As in that paper, the ratio of the galaxies' scale lengths
was fixed using scaling relations drawn from observations of real
galaxies.
In particular, for $\gamma_1=1$ and $\gamma_2=2$, we took
$r_{e,2}/r_{e,1}= (M_2/M_1)^{3/5}$.
The primary galaxy had $N_1=5\times 10^5$ equal-mass particles. 
For the secondary galaxy we took $N_2=(M_2/M_1)N_1=1.66\times 10^5$ so that
all particles had the same mass.
We set the initial orbital parameters for the mergers such that
the separation between the galaxy centers was $\sim 3$ times 
the half-mass radius of the primary  galaxy. The initial orbital 
velocities were assigned in units of the angular
momentum of a circular orbit at the defined separation, 
$\kappa \equiv L/L_{cir}$, with 
$\kappa$=$\left\{0,0.2,0.5,0.8\right\}$.
   
The evolution was followed using the tree code GADGET (Springel et 
al. 2000), a parallel algorithm with continuously-variable time steps 
for each particle. The merger simulations were continued until $\sim$ 1
dynamical time of the primary galaxy after formation 
of the hard black-hole binary, slightly longer than in the simulations
of Paper I.
Integrations were carried out using the HPC10000 
supercomputer at the Rutgers Center for Advanced Information 
Processing and the Cray T3E at the San Diego Supercomputer 
Center. All simulations used 16 processors.

\begin{figure}
\vspace{1.2in}
\plotfiddle{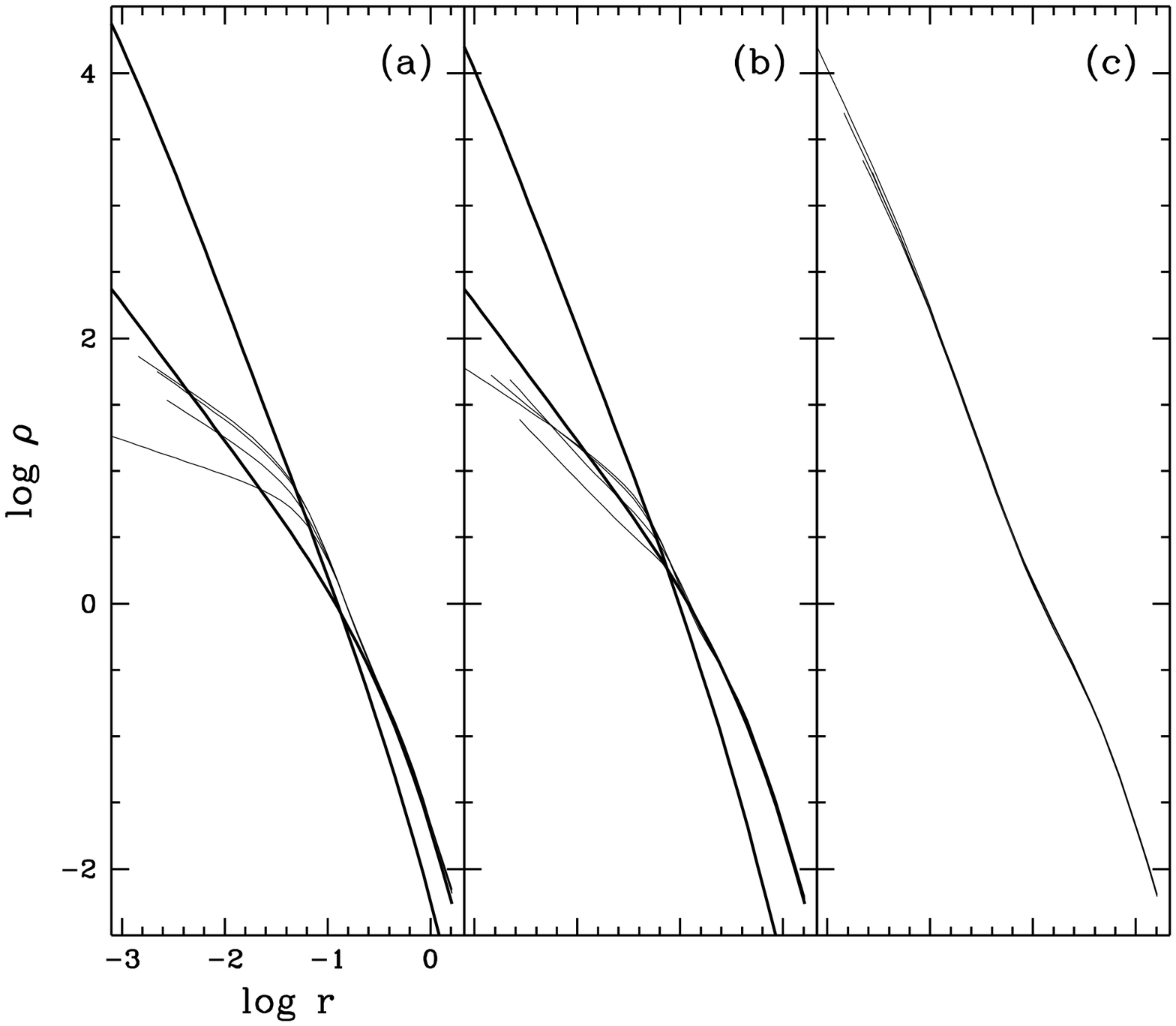}{3.in}{0}{65.}{50.}{-200.}{-50.}
\plotfiddle{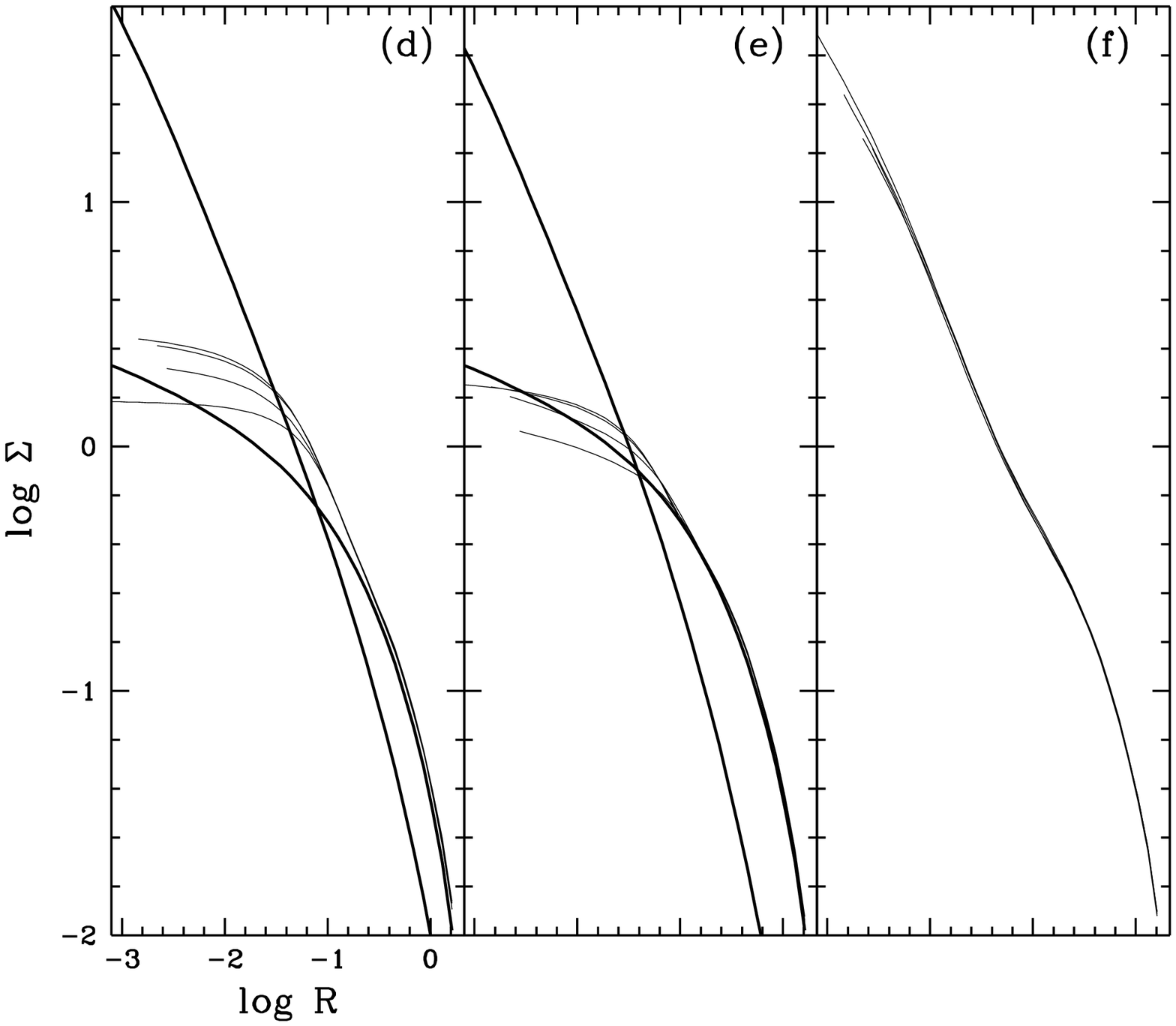}{3.in}{0}{65.}{50.}{-200.}{-75.}
\caption{Final density profiles (a-c) and projected density profiles (d-f)
for all remnants in our simulations. The four thin curves in each
frame correspond to different initial orbits; the thin curve
extending farthest to the left corresponds to  $\kappa$=0.8 and the
shortest to $\kappa$=0. Thick lines are the initial
profiles. In (a) and (d) we show mergers of galaxies with central
black holes and mass ratio $3:1$; in (b) and (e) the mass ratio is $10:1$. 
In (c) and (f) we show the final density profiles for $10:1$
mergers in which nethier galaxy contained a black hole.}
\end{figure}

\section{\bf Results \& Discussion}

Final density profiles for the merger remnants are shown in 
Figure 1, computed from the particle positions using 
the {\tt MAPEL} package (Merritt 1994). 
We also plot final density profiles from the $10:1$ mergers of
Paper I; because those simulations were extended for slightly longer
times here, the central densities have dropped slightly compared
to the values shown in Paper I due to continued ejection of stars by
the black-hole binary.

The central density slopes for all of our simulations are lower
at the final time step than the initial slope of the giant galaxy.
We find $0.5\la\gamma\la 1$ for the $3:1$ mergers and
$\gamma\la 1$ for the $10:1$ mergers.
In projection, all of the remnants exhibit ``cores,'' regions
of nearly constant surface brightness near the center;
the cores are especially noticeable in the $3:1$ remnants.
This is similar to the situation in real elliptical galaxies,
where nonparametric deprojection of the luminosity profiles 
of the ``core'' galaxies reveals
shallow power-law cusps in the space density (Merritt \& Fridman 1995).
The very mild inflections seen in the projected density profiles
of Figure 1 are also consistent with observed profiles.

Based on these simulations, and on the $1:1$ merger simulations of 
Milosavljevi\'c \& Merritt (2001), we propose the following two rules
that relate initial and final density profiles at the centers of
merging galaxies.
1. Mergers between galaxies containing power-law density cusps
produce remnants with power-law cusps.
2. The final density profile is always shallower near the 
center than the initial profile of the more massive galaxy. 
The first ``law'' appears to be true even in mergers without
black holes (Figure 1c,f; Barnes 1999),
while the second ``law'' only holds if the massive galaxy contains a
supermassive black hole.
Central densities might be driven even lower by the continued ejection
of stars by a binary supermassive black hole (Milosavljevi\'c \& 
Merritt 2001).
These ``laws,'' if they can be shown to hold generally for mergers 
between galaxies with supermassive black holes, could explain how the observed 
relation between $M_V$ and $\gamma$ is preserved in the face of 
mergers.

This work was supported by NSF grant AST 00-71099,
NASA grants NAG5-6037 and NAG5-9046, by a fellowship from
the Consejo Nacional de Ciencia y Tecnologia de Mexico, and by 
grant no. MCA00N010N from NCSA.


\begin{references}
\reference Barnes, J. E. 1999, in IAU Symp. 186, Galaxy Interactions
	at Low and High Redshift, ed. J. E. Barnes \& D. B. Sanders
	(Dordrecht: Kluwer), 137 
\reference Dehnen, W. 1993, MNRAS, 265, 250
\reference Ferrarese, L., van den Bosch, F. C., Ford, H. C., Jaffe, W. \&
        O'Connell, R. W. 1994, AJ, 108, 1598
\reference Gebhardt, K. et al. 1996, AJ, 112, 105
\reference Merritt, D. 1994, 
        http://www.physics.rutgers.edu/$\sim$merritt/mapel\_1.html
\reference Merritt, D. \& Ferrarese, L. 2001, MNRAS, 320, L30
\reference Merritt, D. \& Cruz, F. 2001, ApJ, 551, L41 (Paper I)
\reference Merritt, D. \& Fridman, T. 1995, in Fresh Views of 
	Elliptical Galaxies, A.S.P. Conf. Ser. Vol. 86 
	(ed. A. Buzzoni, A. Renzini \& A. Serrano) 22.
\reference Milosavljevi\'c, M., \& Merritt, D. 2001, astro--ph/0103350
\reference Peebles, P. J.. E. 1972, Gen. Rel. Grav., 3, 63
\reference Quinlan, G. D., Hernquist, L. \& Sirgudsson, S. 1995, Apj, 440, 554
\reference Springle, V., Yoshida, W. \& White, S. D. M. 2000,
	   astro-ph/0003162
\reference Tremaine, S. et al. 1994, AJ, 107, 634
\end{references}
\end{document}